# Automated object classification with ClassX

A.A. Suchkov (STScI), T.A. McGlynn, L. Angelini, M.F. Corcoran, S.A. Drake, W.D. Pence, N. White, E.L. Winter (NASA/GSFC), R.J. Hanisch, R.L. White, M. Postman, M.E. Donahue (STScI), F. Genova, F. Ochsenbein, P. Fernique, & S. Derriere (CDS)

**Introduction.** *ClassX* is a project aimed at creating an automated system to classify X-ray sources and is envisaged as a prototype of the Virtual Observatory. As a system, ClassX integrates into a pipeline a network of classifiers and an engine that searches and retrieves for a given target multi-wavelength counterparts from the worldwide data storage media. At the start of the project we identified a number of issues that needed to be addressed to make the implementation of that kind of a system possible. The most fundamental are: (a) classification methods and algorithms, (b) selection and definition of classes (object types), and (c) identification of source counterparts across multi-wavelength data. Their relevance to the project objectives will be seen in the results below as we discuss ClassX classifiers.

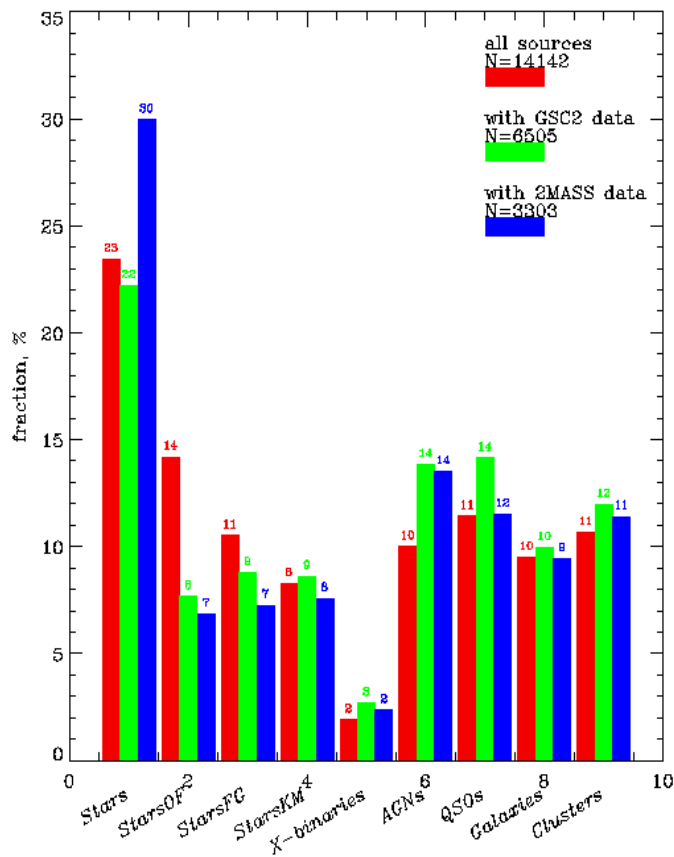

**Fig. 1**. Class distribution of the samples used for training classifiers: (a) all sources classified in the ROSAT WGA catalog, (b) sources with optical (GSC2) counterparts within 30 arcsec, and (c) sources with near-infrared (2MASS) counterparts within 6 arcsec from the respective optical counterpart.

**Classifiers**. We apply machine learning methods to generate classifiers from 'training' data sets, each set being a particular sample of objects with pre-assigned class names that have measured X-ray fluxes and, wherever possible, data from other wavelength bands. In this paper, a classifier is represented by a set of oblique decision trees (DT) induced by a DT generation system OC1. An X-ray source is input into a classifier as a set of X-ray fluxes and possibly data from the optical, infrared, radio, etc. The discussion below includes some results obtained with classifiers trained on the data from the ROSAT WGA, GSC2, and 2MASS catalogs (**Fig. 1**).

**Classifier metrics**. In order to quantify the quality and efficiency of classifiers, we have introduced a number of various metrics. They include: Classifier's *preference*, $P_{ij}$, which is the probability that a class $i$ object will be classified as class $j$ (**Fig. 2**); *affinity*, $A_{ij}$, which is the probability that an object classified as class $i$ does in fact belong to class $j$ (**Fig. 2**); *power*, $S_i$, which is the ratio of the probability that an object classified as class $i$ is indeed class $i$ to the probability that a randomly selected object belongs to class $i$. Very useful characteristics are *completeness*, $C_i = P_{ii}$, and *reliability*, $R_i = A_{ii}$.

**Classifiers' network.** Using different sets of training parameters (attributes), we get different classifiers for the same list of class names (e.g., **Fig. 2**). We integrate them into a network, in which each classifier makes its own class assignment and is optimized for handling different tasks and/or different object types. We envision that, having a set of X-ray sources, a user would generally select a certain classifier to make, for instance, a most *complete* list of candidate QSOs, but a different classifier would be used to make a most *reliable* list of such candidates. Still different classifiers would be selected to make similar lists for other object types. **Fig. 2** suggests that one would prefer the *xray_gsc2* classifier to pick up cluster candidates, while AGNs call for the *xray_only* classifier.

**Training set deficiencies**. A classifier is adversely impacted by source misclassification, counterparts misidentification, data bias, etc. As the training data improve, so do the classifiers. In **Fig. 3**, about 50% of class "Stars" sources (stars without spectral classification) come from the LMC/SMC region. This introduces a coordinate bias that affects classifiers generated from that data. Certain metrics of a classifier can be improved if in the respective training set the stars from the LMC/SMC region are dropped.

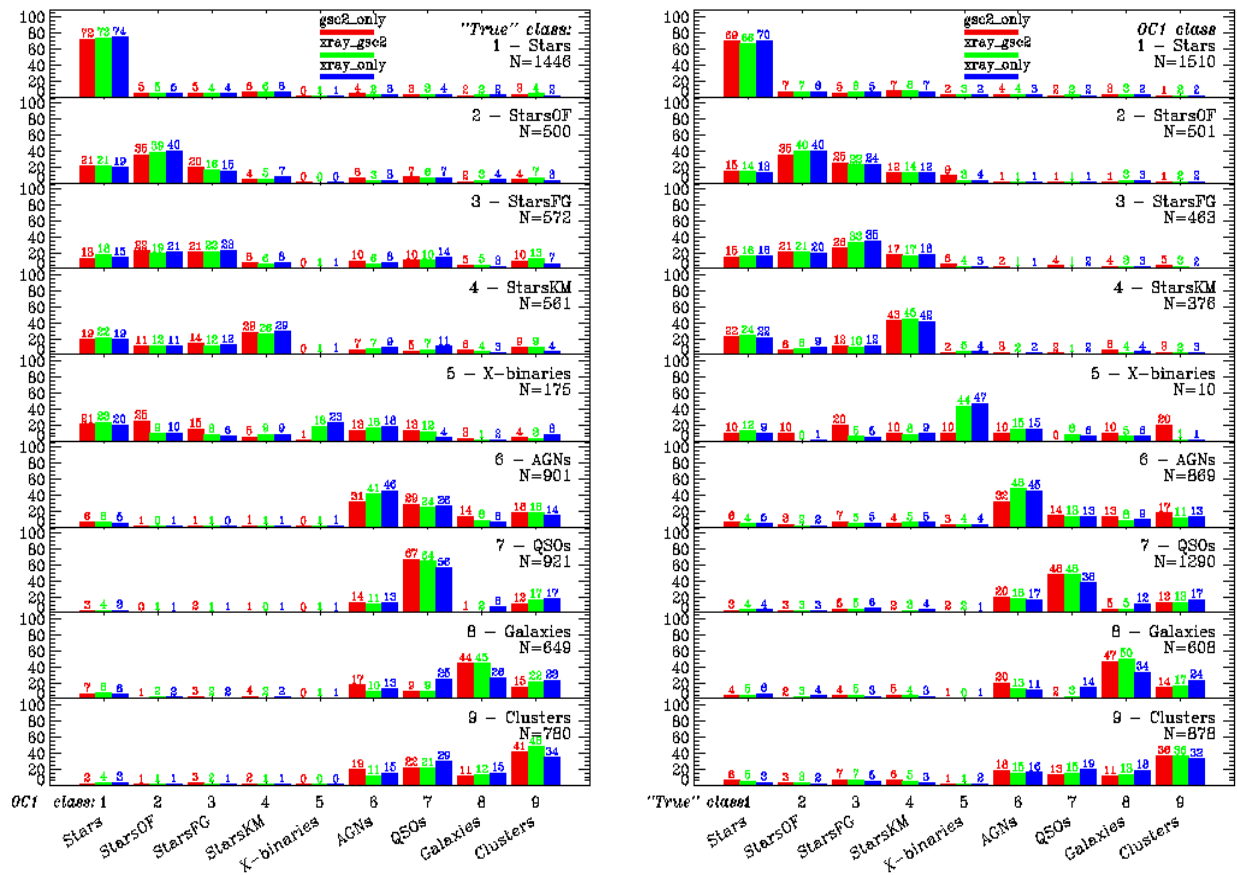

**Fig. 2**. Preference, $P_{ij}$, (left) and affinity, $A_{ij}$, (right) for three classifiers trained using (a) X-ray magnitudes, (b) GSC2 magnitudes, and (c) both GSC2 and X-ray magnitudes along with coordinates and GSC2 "extended" vs. point" source parameter. Notice that the OC1 classifiers separate quite reliably stellar objects from non-stellar ones. At the same time, a confusion between different types of stars or, say, QSO and AGN should be expected because of original misclassification and significant overlap of the respective object types in the parameter space.

**Validating pre-assigned classes with ClassX.** An X-ray source in a training set may have an inappropriate class name and/or get wrong optical or other counterpart. Candidates to have those deficiencies can be identified as a classifier is applied to the training data. In **Fig. 4,** an OC1 classifier is seen to noticeably enhance the contrast between "extended" and "point" sources for StarsKM, QSOs, and Clusters, suggesting that the sources contributing to that enhancement were probably misclassified in the training set. They can further be examined and then reclassified if warranted, which would improve the training set itself.

**Counterpart search strategies with ClassX.** Classifiers trained using optical counterparts proved to be much better if a counterpart is selected as a brightest objects within 30 arcsec as opposed, for instance, to a nearest object within 30 arcsec or a brightest object within 60 arcsec. Thus, classifier validation in ClassX offers a way to find best strategies to search for multi-wavelength counterparts.

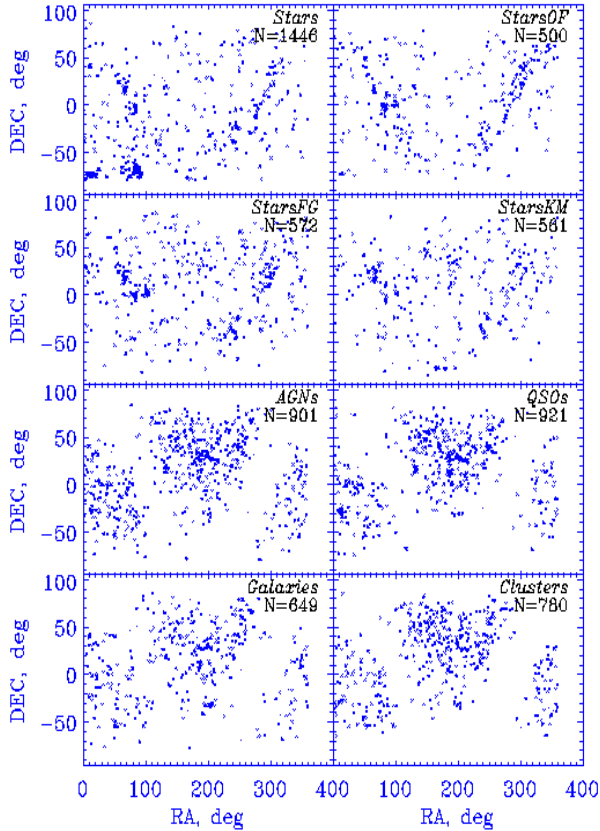

**Fig. 3.** Sky distribution of the ROSAT WGA sources that have GSC2 counterparts reveals that many of the class "Stars" objects are within the LMC/SMC region, which leads to a coordinate bias in classifiers induced from that data.

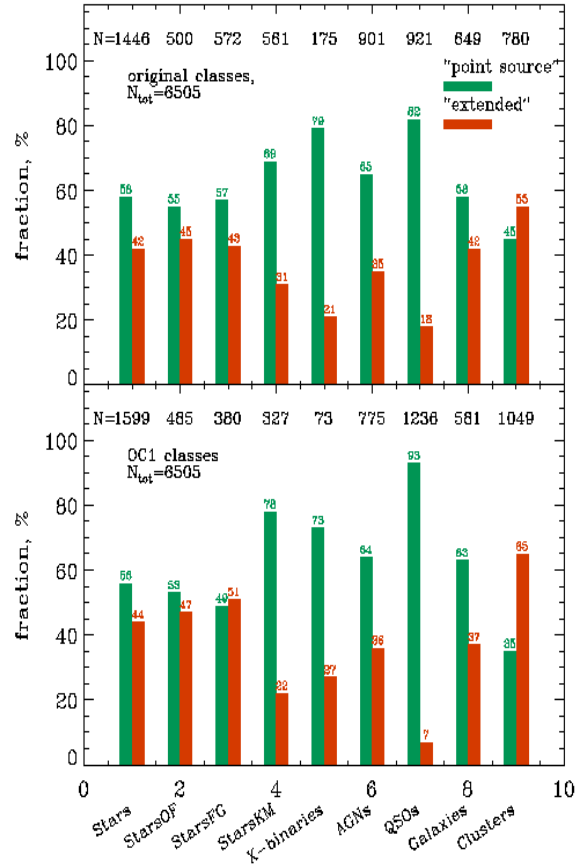

**Fig. 4.** GSC2 source size (3 for "extended", 0 for "pointlike") for the original ("true") and OC1 classifications. Classes from the OC1 classifier are more consistent with the GSC2 source size.

**Class "fuzziness" in ClassX.** A class is rarely a clear-cut notion. Someone's QSO is another person's AGN or a galaxy. ClassX puts this class name "fuzziness" into focus. With ClassX, one can isolate sources with a greater degree of class name ambiguity and look into why their classification in the training set differs from an OC1 classification (e.g., **Fig. 5**).

**ClassX outputs.** A network classifier outputs the class name and the probability that the source belongs to the assigned class. It also outputs the probabilities that the source belongs, in fact, to other classes in the class name list. This allows, for instance, to see how close the source association with various classes in the parameter space is (see **Table 1**)

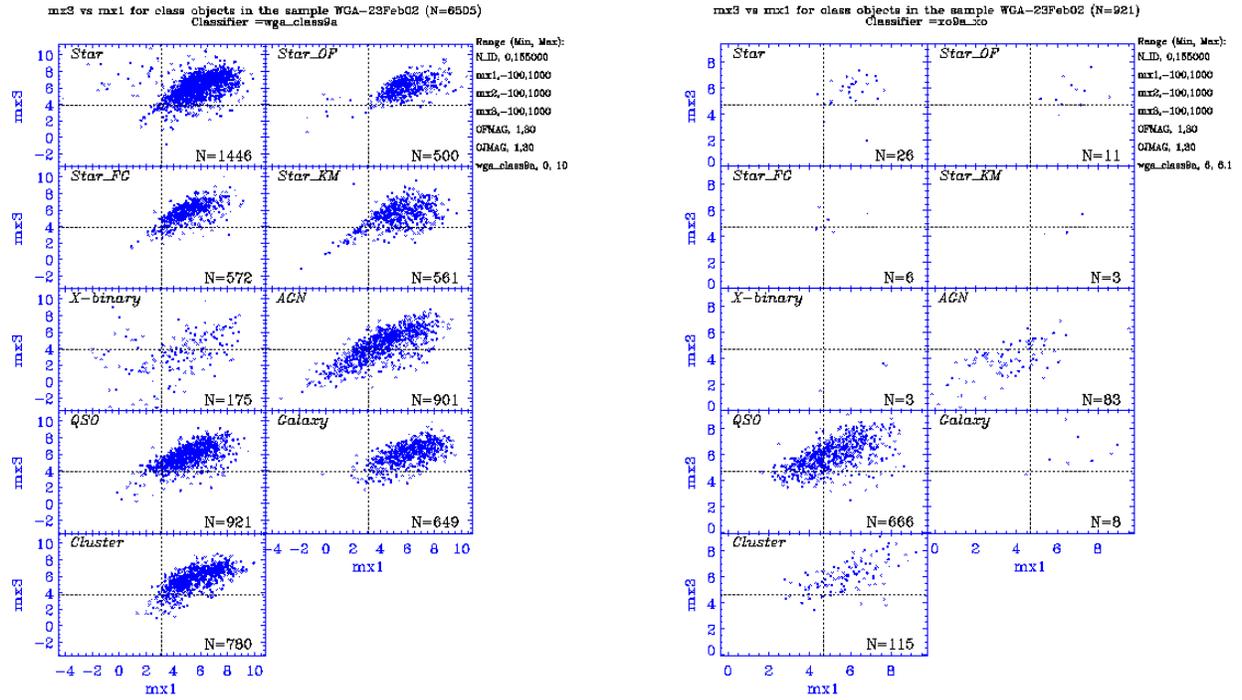

**Fig. 5**. X-ray soft–hard brightness diagram for classes from the WGA catalog (left) and the OC1 classifier *xo9a_xo* (right). In the right, only the WGA class QSO is shown. Most of the brightest and the faintest WGA QSOs have been classified by the *xo9a_xo* as AGN and Cluster, respectively, which partly reflects the class name ambiguity, or "fuzziness", for these sources.

**Table 1:** Illustration of an output from a classifier applied to the training data. P(i) is the probability that a given source belongs to class i, OC1 is the class assigned by the classifier (class coding is the same as in Fig. 2), Input and input name are the class code and class name in the training set ("true" class), respectively. In its vote in the case of the QSO almost equally between Cluste**r** (code 9) and QSO (code 7), marginally favoring Cluster.

| WGA_ID | OC1 | Input | P(1) | P(2) | P(3) | P(4) | P(5) | P(6) | P(7) | P(8) | P(9) | Input name |
|---|---|---|---|---|---|---|---|---|---|---|---|---|
| 1WGA J0458.0-7515 | 1 | 1 | 0.864 | 0.045 | 0.006 | 0.008 | 0.027 | 0.017 | 0.017 | 0.010 | 0.005 | STAR |
| 1WGA J0055.7-7137 | 1 | 1 | 0.882 | 0.045 | 0.005 | 0.002 | 0.027 | 0.014 | 0.015 | 0.008 | 0.003 | STAR |
| 1WGA J1625.8+2646 | 9 | 7 | 0.073 | 0.059 | 0.065 | 0.051 | 0.025 | 0.116 | 0.241 | 0.111 | 0.258 | QSO |

This project is funded through NASA's AISR program (NAG5-11019).